\documentstyle[12pt]{article}

\date{\today}

\def\be{\begin{equation}}
\def\ee{\end{equation}}
\def\bear{\begin{eqnarray}}
\def\eear{\end{eqnarray}}
\def\nn{\nonumber}

\def\wdg{{\wedge}}                              

\newcommand\rep[1]{{\underline{\bf {#1}}}}      

\def\BR{{{\bf R}}}
\newcommand\MS[1]{{{\bf S}^{#1}}}               
\newcommand\MT[1]{{{\bf T}^{#1}}}               

\newcommand\SUSY[1]{{{\cal N}= {#1}}}           

\def\u{{\mu}}
\def\v{{\nu}}

\def\lam{{\lambda}}

\begin{document}

\begin{titlepage}
\titlepage
\rightline{hep-th/0002175, IAS-HEP-00/07, PUPT-1914}
\rightline{February 21, 2000}
\vskip 1cm
\centerline{{\Huge Pinned Branes}}
\centerline{{\Huge and New Non Lorentz Invariant Theories}}
\vskip 1cm
\centerline{
Shoibal Chakravarty$^\clubsuit$,
Keshav Dasgupta$^\diamondsuit$,
Ori J. Ganor$^\clubsuit$ and
Govindan Rajesh$^\diamondsuit$
}
\vskip 0.5cm

\begin{center}
\em  ${}^\clubsuit$Department of Physics, Jadwin Hall \\
Princeton University \\
Princeton, NJ 08544, USA \\
{\it shoibalc@princeton.edu, origa@viper.princeton.edu}
\end{center}

\vskip 0.5cm

\begin{center}
\em  ${}^\diamondsuit$School of Natural Sciences \\
Institute of Advanced Study \\
Princeton, NJ 08540, USA \\
{\it keshav, rajesh@sns.ias.edu}
\end{center}

\vskip 0.5cm

\abstract{
We describe a mechanism for localising branes in ambient space.
When a 3-form flux is turned on in a Taub-NUT space,
an M5-brane gets an effective potential that pins it to
the center of the space.
A similar effect occurs for M2-branes and D-branes with
appropriate fluxes.
In carefully chosen limits of the external parameters, this
leads to new theories that are decoupled from gravity
and appear to break Lorentz invariance.
For example, we predict the existence of a new 5+1D theory
that breaks Lorentz invariance at high-energy and has
a low-energy description of
$N$ tensor multiplets  with ${\cal N}=(1,0)$ supersymmetry.
We also predict a new type of theory that, similarly to the
little-string theory decouples from gravity by a dynamical 
(rather than kinematical) argument.
}
\end{titlepage}
             


\section{Introduction}\label{intro}
The way one usually relates field-theories to branes is
to take the low-energy limit. Thus, taking $M_s\rightarrow\infty$
for $N$ coincident D3-branes leaves the $U(N)$ $\SUSY{4}$ SYM degrees
of freedom only \cite{WitBND}
and taking the $M_p\rightarrow\infty$ limit for $N$ coincident
M5-branes leaves the $(2,0)$ degrees of freedom \cite{StrOPN}.
One can obtain field theories with less supersymmetry by
placing the branes at singularities 
\cite{DouMoo}-\cite{BluInt}\cite{KleWit}.
A larger class of theories is possible if one relaxes the
condition of Lorentz invariance.
One can realize the Lorentz noninvariant theories by placing
the branes in backgrounds that break Lorentz invariance.
The most studied example is Yang-Mills theories
on noncommutative spaces \cite{Connes} obtained by
an appropriate scaling limit of branes in backgrounds with
an NSNS flux \cite{CDS}-\cite{SWNCG}.

The purpose of this paper is to study more configurations
of branes at backgrounds that break Lorentz invariance
and have an interesting low-energy limit.

The construction that we will use is as follows.
Consider a smooth 4D Taub-NUT space (in either M-theory or type-II
string theories) that at infinity behaves as
a circle fibration over the sphere $S^2$ with first Chern class
$c_1 = 1$. To be concrete, let us take M-theory. The Taub-NUT
space is homogeneous in 6+1 directions.
We can turn on a constant 3-form $C$-flux along the circle at infinity and
two of the homogeneous directions, such that the 4-form field-strength
$dC$ is zero at infinity. Because the Taub-NUT circle
shrinks to a point at the origin the 4-form field strength cannot remain
zero throughout the interior of the Taub-NUT space.
In the classical approximation, a solution with
this particular boundary conditions forces a nonzero field-strength
and therefore also affects the metric. The metric is changed in such
a way that a brane that is transverse to the
Taub-NUT space (i.e. parallel to the 6+1 directions)
would have a lower tension if it is at the center of the space.

By tuning the external parameters (the $C$ flux at infinity and
the radius of the Taub-NUT circle at infinity) we can decouple gravity.
In fact, we will suggest two possible limits that decouple gravity.
In the first limit the flux is small
and the low-energy theory appears to be a new kind
of a 5+1D theory with $\SUSY{(1,0)}$ supersymmetry 
that can roughly be described as the $(2,0)$ theory with 
a massive hypermultiplet. There is no contradiction with chirality
of the hypermultiplet in 5+1D because we believe the theory 
(and, in particular, the mass term) breaks Lorentz invariance explicitly.
In the second limit the flux is kept finite and the decoupling argument
is of the same nature as  the dynamical argument presented in \cite{SeiVBR}.

Similar constructions can be repeated with the M2-brane and D-branes.

The paper is organized as follows.
In section (\ref{setting})
we will describe the setting for the constructions
and review the geometry of the Taub-NUT space.
In section (\ref{bps})
we will study the spectrum of BPS excitations of the 
pinned branes and the various energy scales involved.
We will show that the spectrum includes a particle with a very low mass.
In section (\ref{gravity})
we will interpret some of the results of section (2)
from the low-energy supergravity solutions.
In section (\ref{decoup})
we will present the limits of the external parameters
that decouples gravity.
In section (\ref{loweng}) we discuss the low-energy description
of the theories and resolve a puzzle about fermions.
In section (\ref{sevend}) we remove the M5-branes and study the
Taub-NUT space with 3-form flux, $C_3$  on its own.
We analyze the spectrum of BPS states and show that the large
$C_3$ limit can be accompanied with a rescaling of coordinates
that makes the energies of the low-lying BPS states finite.



\section{The setting}\label{setting}

\subsection{Review of the Taub-NUT geometry}
The metric of a KK-monopole is the Taub-NUT metric:

\be\label{tnmet}
ds^2 = R^2 U(dy - A_i dx^i)^2 + U^{-1} (d\vec{x})^2,\qquad
i=1\dots 3,\qquad 0\le y\le 2\pi.
\ee
where,
$$
U = \left(1 + {{R}\over {|\vec{x}|}}\right)^{-1},
$$
and $A_i$ is the gauge field of a monopole centered at the origin.
This metric has a few properties that we will utilize.
\begin{itemize}
\item
It is a circle fibration over $\BR^3$ with the origin
excluded.
\item
The radius of the fiber shrinks to zero as we approach
the origin and becomes a constant $R$ as we approach infinity.
\item
If we restrict to $|\vec{x}| = r$ with constant $r>0$
the circle fibration is equivalent to the Hopf fibration
of $S^1$ over $S^2$.
\item
There is a $U(1)$ isometry $y\rightarrow y+\epsilon$.
It has one fixed point at the origin.
\item
The $U(1)$ isometry acts nontrivially on the tangent space
to the point at the origin.
\end{itemize}

\subsection{Turning on a flux at infinity}
Now suppose that we have a theory of gravity coupled to
a vector field $A_\mu$ and we are looking for a solution
to the equations of motion with boundary conditions such that at
infinity we have a circle fibration over $S^2$ with
$c_1 = 1$ and
there is a constant Wilson line $\int A_y dy = w$ on the circle
at infinity.
If $w\neq 0$, we cannot have $F_{\u\v}=0$ throughout space
because this will force the holonomy $\int A_y dy$ to be constant
contradicting the fact that the circle shrinks to zero at $\vec{x}=0$.
Thus, we expect that the solution with the above boundary conditions
will have a nonzero field-strength near the center of the Taub-NUT
space. We also expect the Taub-NUT metric to change, as a consequence.

Now let us turn to the setting in our case.
We take a Taub-NUT space in M-theory or one of the type-II
string theories and we turn on a tensor field at infinity.
In M-theory we take the Taub-NUT circle
direction to be the $7^{th}$ and
let $0\dots 6$ be directions perpendicular to the Taub-NUT space.
We can then  turn on $C_{167}$ at infinity.
In a low energy limit, we will see that there exists a solution
with this kind of boundary condition (i.e. being a Hopf fibration
at infinity and having the constant $C_{167}$ flux).
It is also very plausible that for any value of $C_{167}$ there
exists a background of M-theory with these boundary conditions.
Next, we add M5-branes  along directions
$0\dots 5$ and ask whether there is a limit of $R$
and $C_{167}$ for which we obtain a theory that is decoupled
from gravity.



\def\wQ{{\widetilde{Q}}}
\def\wq{{\widetilde{q}}}
\def\wR{{\widetilde{R}}}

\section{BPS states}\label{bps}
We wish to study the dynamics of $N$ M5-branes at the center
of a Taub-NUT space with a 3-form field that is constant at infinity
and has one direction along the Taub-NUT circle.
In order to understand the dynamics and the relevant energy scales
of the system we will study various BPS states and fluxes in this
theory.

\subsection{Embedding in M-theory on $T^7$}
We would like to use the formula for masses of BPS particles
in M-theory on $T^7$ (see \cite{WitVAR,OPRev}).
We will start with all fluxes turned off and take $T^7$ in the form
of a product of circles of radii $R_1\dots R_7$.
We will denote by $M_p$ the 11-dimensional Planck scale.
We take the Taub-NUT circle to be $R_7$. We let the Taub-NUT wrap
directions $1\dots 6$. Its mass is:
$$
M_{TN} = M_p^9 V R_7,\qquad V\equiv R_1\dots R_7.
$$
The mass of the M5-brane is:
$$
M_{M5} = M_p^6 R_1 R_2 R_3 R_4 R_5  = M_p^6 V R_6^{-1} R_7^{-1}.
$$
There are $56$ $U(1)$ charges in the low-energy description of
M-theory on $T^7$. 
We will denote them by:
$$
Q^i, \wQ_i, Q_{ij}, \wQ^{ij}.
$$
They correspond to BPS particles with  masses:
$$
R_i^{-1},\, M_p^9 V R_i,\, M_p^3 R_i R_j,\, 
M_p^6 V R_i^{-1} R_j^{-1}.
$$
These are KK-particles, KK-monopoles, M2-branes and M5-branes 
respectively.
Now let us turn on some flux $C_{mnp}$ ($1\le m<n<p\le 7$) and
let us assume, for simplicity,
that $C_{mnp}\neq 0$ for only one set of indices $m,n,p$.
Let us consider a BPS-state (that preserves some fraction of
the SUSY) with integer charges $Q^i,\wQ_i, Q_{ij}, \wQ^{ij}$.
 From this vector one can construct an $8\times 8$ complex 
anti-symmetric central
charge matrix $Z_{ab}=-Z_{ba}$ ($a,b=1\dots 8$) that is linear
in the $Q$'s. The procedure is as follows \cite{WitVAR}.
Let us define the periodic variable:
$$
\phi^{mnp} \equiv C_{mnp} R_m R_n R_p,\qquad
 \phi^{mnp}\sim \phi^{mnp}+1.
$$
Next, one defines:\footnote{We wish to thank the anonymous referee for
pointing out a typo in the first line.
This corrects two formulas in section (\ref{sevend}).}
\bear
q^i &\equiv& Q^i +\phi^{ijp} Q_{jp},\nn\\ 
\wq_i &\equiv& \wQ_i,\nn\\
\wq^{ij} &\equiv& \wQ^{ij} + \phi^{ijp}\wQ_p,\nn\\
q_{ij} &\equiv& Q_{ij} -\epsilon_{ijklmnp} \wQ^{kl}\phi^{mnp}.\nn
\eear
Here we have used the assumption that $\phi_{mnp}\neq 0$ for only
one set of indices $m,n,p$ (that we are going to take to be $1,6,7$ later
on). Otherwise, we will also have terms that are quadratic and cubic
in $\phi$.
One way to think about these equations is that the presence of the
fractional $\phi_{mnp}$ creates effective fractional
charges. For example, there is an effective fractional membrane charge
if there is an M5-brane with a $C$-field turned on.

The central charge matrix is now given by:
$$
Z_{ab} = \sum_{1\le m<n\le 7} (M_p^3 R_m R_n q_{mn}
        +i M_p^6 V R_m^{-1} R_n^{-1} \wq^{mn})\Gamma^{mn}_{ab}
       +\sum_{m=1}^7 (M_p^9 V R_m \wq_m
        +i R_m^{-1} q^m)\Gamma^{m8}_{ab}.
$$
Here we have used the anti-symmetric
$$
\Gamma^{pq}_{ab}=-\Gamma^{qp}_{ab} = -\Gamma^{pq}_{ba},\qquad
p,q=1\dots 8,\qquad a,b=1\dots 8
$$
which are the generators of $SO(8)$ in the spinor representation
$\rep{8}_s$.
The BPS bound from the matrix $Z$ is that the mass squared of a state
with given charge should be at least the maximal eigenvalue
of $Z^\dagger Z$.

We shall now apply this formula to various BPS states in the theory.

\subsection{The 5-brane tension}
How much energy does it cost to separate the 5-brane
from the Taub-NUT space?

We set $C_{167}$ to a nonzero value and set $\wQ_7 = 1$
and $\wQ^{67} = N$. We also set $\phi^{167} = C_{167} R_1 R_6 R_7$.
The central charge matrix is given by:
$$
Z = M_p^9 V R_7 \Gamma^{78} 
  + i N M_p^6 V R_6^{-1} R_7^{-1}\Gamma^{67}
  + i C_{167} M_p^6 V R_7\Gamma^{16}.
$$
The maximal eigenvalue of this matrix is:
$$
M_p^6 V \sqrt{(M_p^3 R_7 + N R_6^{-1} R_7^{-1})^2 + C_{167}^2 R_7^2}
$$
We compare this to the separate masses of the Taub-NUT alone
and the M5-brane alone.
The sum of the masses is:
$$
M_p^6 V 
\left(\sqrt{M_p^6 R_7^2 + C_{167}^2 R_7^2}  +N R_6^{-1} R_7^{-1}\right)
$$
Thus, the bound state energy is:
$$
M = M_p^6 V
\left(\sqrt{M_p^6 R_7^2 + C_{167}^2 R_7^2} +N R_6^{-1} R_7^{-1}
 - \sqrt{(M_p^3 R_7 + N R_6^{-1} R_7^{-1})^2 + C_{167}^2 R_7^2}\right)
$$
Let us take the limit $M_p^3 R_6 R_7^2\rightarrow\infty$.
We obtain ($C\equiv M_p^{-3} C_{167}$):
$$
M\approx N M_p^6 V R_6^{-1} R_7^{-1} 
 \left(1 - {1\over {\sqrt{1 + C^2}}}\right).
$$
Thus, the tension of each 5-brane effectively decreases
by $\sqrt{1+C^2}$ when it is bound to the Taub-NUT.

\subsection{Momentum States}
The result above can be interpreted simply as a rescaling
of the metric $g_{11}$ in the $1^{st}$ direction.
To see this, let us compare the energy of a KK-particle
in the $1^{st}$ direction to the energy of a KK-particle
with momentum in the $2^{nd}$ direction.

For momentum in the $1^{st}$ direction we find:
\bear
Z &=& M_p^9 V R_7 \Gamma^{78} 
  + i N M_p^6 V R_6^{-1} R_7^{-1}\Gamma^{67}
  + i C_{167} M_p^6 V R_7\Gamma^{16}
  + i k R_1^{-1} \Gamma^{18}
\nn\\
&\equiv& x\Gamma^{78}+i y\Gamma^{67} +iz\Gamma^{16} + i v\Gamma^{18}.
\nn
\eear
The mass is:
$$
\sqrt{(x+y+v)^2 + z^2}
-\sqrt{(x+y)^2 + z^2}\longrightarrow
{{x}\over {\sqrt{x^2+z^2}}} v
$$
The last result is in the limit $x,z\gg y\gg v$.
Thus the energy of a massless particle with $k$ units of momentum
in the $1^{st}$ direction is:
$$
{k\over {\sqrt{1+C^2}}} R_1^{-1}.
$$
This suggests that we define:
$$
\wR_1 \equiv \sqrt{1+C^2} R_1
$$
The KK-mass is then $\wR_1^{-1}$.
For particles with momentum in the $2^{nd}$ direction we find:
\bear
Z &=& M_p^9 V R_7 \Gamma^{78} 
  + i N M_p^6 V R_6^{-1} R_7^{-1}\Gamma^{67}
  + i C_{167} M_p^6 V R_7\Gamma^{16}
  + i k R_2^{-1} \Gamma^{28}
\nn\\
&\equiv& x\Gamma^{78}+i y\Gamma^{67} +iz\Gamma^{16} + i v\Gamma^{28}.
\nn
\eear
The mass is then just $k R_2^{-1}$.

\subsection{Massive Particles}
As we have seen, the $N$ M5-branes are stuck at the center of
the Taub-NUT space where their tension is minimal.
Intuitively, this suggests that small fluctuations of the world-volume
of the M5-branes are described by a massive field.
A Taub-NUT space has a $U(1)$ isometry.
When $C_{167}=0$, a fluctuation of the M5-brane in the
Taub-NUT directions is charged under that $U(1)$ because at
the center of the Taub-NUT geometry the $U(1)$ is embedded in
the local $SO(4)$ isometry of the tangent space.

Thus, we should check what is the mass of a BPS state with $U(1)$
charge. We therefore set $Q^7 = k$ and calculate:
$$
Z = (M_p^9 V R_7 +i k  R_7^{-1})\Gamma^{78} 
  + i N M_p^6 V R_6^{-1} R_7^{-1}\Gamma^{67}
  + i M_p^6 C_{167} V R_7\Gamma^{16}.
$$
Now the maximal eigenvalue is:
$$
M_p^6 V \sqrt{(M_p^3 R_7 + N R_6^{-1} R_7^{-1})^2 + 
      (C_{167} R_7 + k M_p^{-6} V^{-1} R_7^{-1})^2}
$$
The energy of the excitation is therefore:
$$
M_p^6 V \left(\sqrt{(M_p^3 R_7 + N R_6^{-1} R_7^{-1})^2 
       + (C_{167} R_7 + k M_p^{-6} V^{-1} R_7^{-1})^2}
     -\sqrt{(M_p^3 R_7 + N R_6^{-1} R_7^{-1})^2 + C_{167}^2 R_7^2}\right).
$$
In the limit $M_p^3 R_6 R_7^2\rightarrow\infty$ this becomes:
$$
k {{C}\over {\sqrt{1+C^2}}} R_7^{-1}.
$$

\subsection{Tensor fluxes}
We can also calculate the energy of fluxes of the anti-self-dual
2-form field that is part of the low-energy tensor multiplet
of an  M5-brane.
Because $C_{167}$ explicitly breaks Lorentz invariance,
we should discuss fluxes in various directions separately.
We set $Q_{ij} = N_{ij}$ for $N_{ij}$ units of tensor flux in the direction
$i,j$. If $T_{ijk}$ is the anti-self-dual 3-form field-strength
on the M5-brane then $N_{ij} = 2\pi T_{ij0} R_i R_j$ for a single M5-brane.

Let us first set only $N_{23}\neq 0$.
The central charge matrix is:
\bear
Z &=& M_p^9 V R_7 \Gamma^{78} 
  + i N M_p^6 V R_6^{-1} R_7^{-1}\Gamma^{67}
  + i C_{167} M_p^6 V R_7\Gamma^{16}
  + M_p^3 N_{23} R_2 R_3\Gamma^{23}
\nn\\
&\equiv& x\Gamma^{78}+i y\Gamma^{67} +iz\Gamma^{16} + u\Gamma^{23}.
\nn
\eear
The BPS bound is:
$$
\sqrt{\left(x+\sqrt{u^2 + y^2}\right)^2 + z^2}
$$
In the limit $M_p^3 R_6 R_7^2\rightarrow\infty$,
$x,z\gg y,u$ we obtain the energy
of the flux:
\bear
E_{23} &=&
\sqrt{\left(x+\sqrt{u^2 + y^2}\right)^2 + z^2}
- \sqrt{(x+y)^2 + z^2} = {{\sqrt{u^2 + y^2} - u}\over {\sqrt{x^2+z^2}}}
\nn\\
&=&
{{\sqrt{1+M_p^{-6} N^{-2} N_{23}^2 R_1^{-2} R_4^{-2} R_5^{-2}}-1}\over 
    {\sqrt{1+C^2}}}M_p^6 N R_1 R_2 R_3 R_4 R_5
\nn\\
&\longrightarrow&
  {{N_{23}^2 R_2 R_3}\over {2\sqrt{1+C^2} N R_1 R_4 R_5}}
= {{N_{23}^2 R_2 R_3}\over {2 N \wR_1 R_4 R_5}}.
\nn
\eear
The last line is in the limit $M_p R_i\gg 1$ for $i=1,4,5$.
Next we check a flux with one index in the direction of $C_{167}$.
Let us take $N_{12}\neq 0$.
We find:
\bear
Z &=& M_p^9 V R_7 \Gamma^{78} 
  + i N M_p^6 V R_6^{-1} R_7^{-1}\Gamma^{67}
  + i C_{167} M_p^6 V R_7\Gamma^{16}
  + M_p^3 N_{12} R_1 R_2\Gamma^{12}
\nn\\ 
&\equiv& x\Gamma^{78}+i y\Gamma^{67} +iz\Gamma^{16} + v\Gamma^{12}.
\nn
\eear
\bear
E_{12} &=&
\sqrt{v^2 + x^2 + y^2 + z^2 +2\sqrt{v^2 x^2 +x^2 y^2 + v^2 z^2}}
 - \sqrt{(x + y)^2 + z^2}
\nn\\ 
&\longrightarrow&
{{\sqrt{v^2(1+C^2)+y^2} - y}\over {1+C^2}}
\longrightarrow_{y\gg v}  {{v^2\sqrt{1+C^2}}\over {2 y}}
\nn\\
&=&
  {{N_{12}^2 \sqrt{1+C^2} R_1 R_2}\over {2 N R_3 R_4 R_5}}
= {{N_{12}^2 \wR_1 R_2}\over {2 N R_3 R_4 R_5}}.
\nn
\eear
We see that these results support the claim that we have
to rescale the first coordinate by a factor of $\sqrt{1+C^2}$.

\subsection{Strings}
So far we described the excitations of a possibly free theory.
Now we would like to take the number of M5-branes to be
 $N=2$ and separate the M5-branes along
the $6^{th}$ direction. Let the separation be $\phi R_6$.
We expect to find strings, made by M2-branes stretched between
the M5-branes, with a tension proportional to $\phi$.
We also have to establish the coefficient of the kinetic term,
$(\partial\phi)^2$, in the low energy effective action.
We can do that by taking $R_2\gg R_6$ and calculate the energy
of and M5-brane that wraps the diagonal of the $2-6$ directions.
This is described by $\phi(x_2) = l x_2 R_2^{-1} R_6$.
The corresponding central charge matrix is:
\bear
Z &=& M_p^9 V R_7 \Gamma^{78} 
  + i N M_p^6 V R_6^{-1} R_7^{-1}\Gamma^{67}
  + i l M_p^6 V R_2^{-1} R_7^{-1}\Gamma^{27}
  + i C_{167} M_p^6 V R_7\Gamma^{16}
\nn\\ 
&\equiv& x\Gamma^{78}+i y\Gamma^{67} +iz\Gamma^{16} + i w\Gamma^{27}.
\nn
\eear

The energy is:
$$
E = \sqrt{x^2 + y^2 + z^2 + w^2 + 2\sqrt{x^2 y^2 + x^2 w^2 + w^2 z^2}}
 - \sqrt{(x+y)^2 + z^2}
\approx {{\sqrt{x^2+z^2}}\over {2xy}} w^2
$$
This is:
$$
\sqrt{1+C^2} {{M_p^6 V R_6}\over {2N R_7}}l^2
= {{M_p^6 \wR_1 R_2 R_3 R_4 R_5}\over {2N}}
\left({{R_6 l }\over {R_2}}\right)^2.
$$
We compare this to:
$$
{1\over {2N}}\wR_1\cdots R_5 (\partial\phi)^2
$$
and find that:
$$
\varphi\equiv M_p^3 \phi
$$
has the normalized kinetic energy.

To calculate the tension of the string we gradually change $\phi$ from
$0$ to $2\pi$. 
When it is $2\pi$ we calculate the mass of $k$ strings stretched on
the $2^{nd}$ direction:
\bear
Z &=& M_p^9 V R_7 \Gamma^{78} 
  + i N M_p^6 V R_6^{-1} R_7^{-1}\Gamma^{67}
  + k M_p^3 R_2 R_6\Gamma^{26}
  + i C_{167} M_p^6 V R_7\Gamma^{16}
\nn\\
&\equiv& x\Gamma^{78}+i y\Gamma^{67} +iz\Gamma^{16} + w\Gamma^{26}.
\nn
\eear
We find $E=w$.
The energy of the string in the limit $R_6\rightarrow\infty$ is thus:
$$
M_p^3 R_2 \phi.
$$
The tension of the string wrapped in a direction not including
the $1^{st}$ is therefore $T\equiv \varphi$.

For the mass of strings stretched in the $1^{st}$ direction we calculate:
\bear
Z &=& M_p^9 V R_7 \Gamma^{78} 
  + i N M_p^6 V R_6^{-1} R_7^{-1}\Gamma^{67}
  + k M_p^3 R_1 R_6\Gamma^{16}
  + i C_{167} M_p^6 V R_7\Gamma^{16}
\nn\\
&\equiv& x\Gamma^{78}+i y\Gamma^{67} +(w+iz)\Gamma^{16}.
\nn
\eear
$$
E = \sqrt{(x+y+w)^2+z^2} - \sqrt{(x+y)^2 + z^2}
 \approx {{x+y}\over {\sqrt{(x+y)^2 + z^2}}} w
$$
The energy of the string in the limit $R_6\rightarrow\infty$ is thus:
$$
{1\over {\sqrt{1+C^2}}}M_p^3 R_1 \phi
= {1\over {1+C^2}} M_p^3 \wR_1\phi.
$$

We note that
in these calculations we assume that the metric on the 
$\varphi$-moduli space is constant. In 5+1D, supersymmetry
implies that the metric on the tensor-multiplet moduli space
is flat. However, in our case we can only use the $SO(4,1)$
subgroup of the $SO(5,1)$ Lorentz group. So, in principle
there can be a nontrivial metric on the moduli space
\cite{SeiFDS}.

\subsection{Summary}
Based on the BPS analysis we found the following facts.

\begin{itemize}
\item
We have to rescale the metric in the $1^{st}$ direction
so that $\wR_1 = \sqrt{1+C^2} R_1$ where $C = M_p^{-3} C_{167}$.
This way the energy of massless particles with momentum $p$
is given by the Lorentz invariant expression $|p|$.
\item
The tension of each M5-brane is smaller by a factor of
$\sqrt{1+C^2}$ when it is at the center of the Taub-NUT
(relative to infinity).
\item
There appears to be a massive particle in the spectrum with
mass
\be\label{bpsmass}
m_0 \equiv {C\over {\sqrt{1+C^2}}} R_7^{-1}.
\ee
\item
The energy of tensor fluxes is as it should be if expressed
in terms of $\wR_1$.
Tensor fluxes in direction $1,2$ have energy:
$$
E_{12}
= {{N_{12}^2 \wR_1 R_2}\over {2 N R_3 R_4 R_5}},
$$
while tensor fluxes in direction $2,3$ have energy:
$$
E_{23} = {{N_{23}^2 R_2 R_3}\over {2 N \wR_1 R_4 R_5}}.
$$
\item
When the M5-branes are separated there appear to be strings
in the spectrum.
The tension of the strings is proportional to the separation.
The tension seems to be smaller by a factor of $(1+C^2)$
for strings stretched in the $1^{st}$ direction relative
to strings stretched in the directions $2\dots 5$.
\end{itemize}



\section{Gravity solutions}\label{gravity}
We have seen in section (\ref{bps}) that 
a Taub-NUT space with nonzero boundary
conditions for the 3-form field along the circle at infinity
creates a potential that pins M5-branes to the origin.
We have also seen that some excitations of the M5-brane
become massive.
In this section we will explain the mechanism that is responsible
for these effects. For that purpose we will describe the classical
supergravity solution that corresponds to this Taub-NUT space.
The solution is a good approximation when the curvature and
field strength are small
and that is true for $M_p R_7\gg 1$  and $M_p^{-3} C_{167} \ll 1$.
We will also describe the solution of the Taub-NUT space
with $N$ M5-branes at the center. This solution is a good
approximation either when $N$ is large or $N=0$.

\subsection{The solution}\label{grasol}
We start with a configuration of $Q_5$  D5-branes and $Q_3$ D3-branes
in type-IIB theory oriented along
($x_0, x_2, x_3, x_4, x_5, x_6$) and ($x_0, x_2, x_3, x_7$)
respectively. We have kept $x_1$ to be the $11^{th}$
direction so as to remain
consistent with the notations of the previous sections.

Under an S-duality this will become a system of $Q_5$  NS5-branes
and $Q_3$ D3-branes. We define the following terms:
$$
H_3= 1 + Q_3/r,\qquad
H_5=1+Q_5/r
$$
and $r= \sqrt{(x^8)^2+(x^9)^2+(x^{10})^2}$. 
$Q_i$ depends on the number $N_i$ of D-branes and also on $M_p$.
We will
use the explicit form of the $Q_i$ later. The metric for the NS5-D3 
configuration  is (see \cite{Tseytlin,Tsey,Gaunt}):
$$
ds^2 = H_3^{-1/2}ds^2_{023}+H_3^{1/2}ds^2_{456}
+ H_5 H_3^{-1/2}ds^2_7+ H_5H_3^{1/2}ds^2_{89,10}
$$
After a series of T-dualities in the $4^{th}$ and $5^{th}$
directions,
we get a configuration of $Q_5$ NS5-branes and $Q_3$ D5 oriented along
($x_0, x_2, x_3, x_4, x_5, x_6$) and ($x_0, x_2, x_3, x_4, x_5, x_7$)
respectively with metric:
$$
ds^2 = H_3^{-1/2}ds^2_{02345}+ H_3^{1/2}ds^2_6 
+ H_5H_3^{-1/2}ds^2_7+H_5H_3^{1/2}ds^2_{89,10}.
$$
Basically this is our starting configuration.
The whole chain of dualities was
done only to calculate the metric for this configuration. The directions 
$x^6,x^7$ are on a square torus. To go to an inclined torus we make the 
following transformations:
\bear
x^6 &=& y^6\sec\theta + y^7\sin\theta,
\nn\\
x^7 &=& y^7\cos\theta.
\nn
\eear
$y^i$ are the new coordinates. Observe that $x^j=y^j$ for $j \not = 6,7$. 
Therefore only the 6,7 part of the metric will undergo some change,
and it will look like:
\bear
ds^2 &=& H_3^{1/2} (dy^6\sec\theta + dy^7\sin\theta)^2 + H_5H_3^{-1/2} 
(dy^7\cos\theta)^2
\nn\\
&=& H_3^{1/2} (dy^6\sec\theta)^2 +(H_3^{1/2}\sin^2\theta + H_5H_3^{-1/2} 
\cos^2\theta) (dy^7)^2 + 2 H_3^{1/2}\tan\theta dy^6 dy^7
\nn
\eear
The rest of the components of the metric will remain the same.
Now under a T-duality 
along $y^7$ we get a ($Q_5$-centered) Taub-NUT space
and $Q_3$ D4-branes. The 
Taub-NUT space has a nontrivial metric along 
($y_7, y_8, y_9, y_{10}$) and the D4-branes are oriented along
($y_2, y_3, y_4, y_5$).

The metric for this configuration is:
$$
ds^2 = H_3^{-1/2} ds^2_{02345} + hH_5(dy^6)^2 + h (dy^7 + B_{7i}dy^i)^2 
+ H_5H_3^{1/2} ds^2_{89,10}
$$
where $h^{-1} = H_3^{1/2}\sin^2\theta + H_5 H_3^{-1/2}\cos^2\theta$
and $i = 8,9,10$.
Observe that the KK gauge field $A_i = B_{7i}$ and $B_{7i}$ 
comes from the NS5 brane. The Taub-NUT circle is along $y_7$. 

There is also an antisymmetric two form background coming from the
inclination of the torus. It is given by:
$$
B^{(NS)} = h H_3^{1/2}\tan\theta\, dy^6\wdg (dy^7 + B_{7i} dy^i)
$$
The coefficient goes to a constant $T=\tan\theta$ at infinity.
The dilaton behaves as:
$$
e^{2\phi} =  hH_5/H_3.
$$
  The string coupling constant, $g$, has been set to one. 

We now lift this configuration to M-theory.
The M-theory direction is $x^1$. The 
various components of the metric are ($a,b$ are 10-dimensional indices):

$$
G_{ab}= (h H_5/H_3)^{-1/3} g_{ab},\qquad
G_{a1}=0,\qquad
G_{11} = (h H_5/H_3)^{2/3}
$$
and the three form background is:
$$
C_3 = M_p^3 h H_3^{1/2}\tan\theta\,
   dx_1\wdg dy^6\wdg (dy^7 + B_{7i} dy^i)
$$
At infinity, the value of the 3-form flux becomes
$M_p^3 C \equiv M_p^3 \tan\theta$.

We can still subtract  from $C_3$ a constant because this
does not affect the field-strength. We do it so as to fix the boundary
condition $C_3=0$ at $r=0$. This is because the radius
of the circle in the $7^{th}$ direction shrinks to zero and
and if $C_3(0)\neq 0$ we would get a singularity in
the field-strength at the origin.
We calculate the constant piece to be:
$$
{{M_p^3 Q_3\tan\theta}\over {Q_3 \sin^2\theta + Q_5\cos^2\theta}}.
$$
Now let us study the background geometry alone, setting the number
$Q_3$ of M5-branes, to zero.
We obtain the metric:
\bear
ds^2 &=&
(hH_5/H_3)^{2/3} dx_1^2 
 +(hH_5)^{-1/3} H_3^{-1/6}(dx_0^2 +dx_2^2 +dx_3^2 +dx_4^2 +dx_5^2)
  + (h H_5)^{2/3} H_3^{1/3} (dy^6)^2 
\nn\\ &&
+ h^{2/3}(H_5/H_3)^{-1/3} (dx_7 + B_{7i}dx_i)^2 
+ h^{-1/3} H_5^{2/3} H_3^{5/6} (dx_8^2 +dx_9^2 +dx_{10}^2)
\nn\\
&=&
(hH_5)^{2/3} dx_1^2 
 +(hH_5)^{-1/3} (dx_0^2 +dx_2^2 +dx_3^2 +dx_4^2 +dx_5^2)
\nn\\ &&
  + (hH_5)^{2/3}  (dy^6)^2 + h^{2/3}H_5^{-1/3} (dx_7 + B_{7i}dx_i)^2 
+ h^{-1/3}H_5^{2/3}  (dx_8^2 +dx_9^2 +dx_{10}^2)
\nn
\eear
where:
$$
H_5 = 1 + {R\over { r}},\qquad h^{-1} = \sin^2\theta + H_5\cos^2\theta
 = 1 + {{R \cos^2\theta}\over {r}}.
$$
We also find:
\bear
M_p^{-3} C_3
 &=& h (H_3)^{1/2}\tan\theta\,
dx_1\wdg dy^6\wdg (dx_7 + B_{7i} dx_i)
\nn\\ &=&
\left(
1 + {{R \cos^2\theta}\over { r}}
\right)^{-1}\tan\theta\,
 dx_1\wdg dy^6\wdg (dx_7 + B_{7i} dx_i)
\label{cbackg}
\eear
as $r\rightarrow\infty$ we see that
$$
M_p^{-3}C_3(\infty) = \tan\theta\,
dx_1\wdg dy^6\wdg (dx_7 + B_{7i} dx_i)
\equiv
C
dx_1\wdg dy^6\wdg (dx_7 + B_{7i} dx_i)
$$

\subsection{Small $r$}\label{smallr}
For $r \to 0$ we can still trust our supergravity solution as long as
the background value of the three form potential is sufficiently small.
In this limit $h^{-1} = H_5/(1+C^2)$ in the absence of M5 brane.
$H_5$, on the other hand, goes as $R r^{-1}$ for small $r$.
The metric as seen by
the M5 brane is nonsingular and behaves as follows:
\bear
ds^2 &=& 
(1+C^2)dx_1^2 + dx^2_{02345} +(1+C^2)dx_6^2 
\nn\\
&+& (1+C^2){r\over R} (dy_7 + A_idy_i)^2
+{R\over r}(dr^2 +r^2d\Omega^2_{8910})
\label{smrmet}
\eear
We have scaled the coordinates by $(1+C^2)^{1/6}$ to get the above metric.
Here $y_7$ and the angular variables
$\Omega_{8,9,10}$ can be taken to parameterize
an ${\bf S}^3$. If we change variables to $r=u^2$
we see that the $r,x_7,\Omega_{8,9,10}$ parameterize
a smooth 4-dimensional point at $r=0$, as in the ordinary 
Taub-NUT space. 
As $r\rightarrow 0$, the field $C_3$ 
behaves as:
\be\label{Cthree}
C_3\rightarrow 
 {{C}{(1+C^2)}} M_p^3 (r/R)\,
 dx_1\wdg dy^6\wdg (dx_7 + B_{7i} dx_i)
\ee
It is easy to check that the field strength, $F_4 = dC_3$,
has a finite magnitude as $r\rightarrow 0$.

\subsection{The pinning potential}
To calculate the potential we have to compute 
$\sqrt{\det G}$ along the M5-brane directions.
This is given by the following expression:
$$
\det G = G_{00}G_{11}\cdots G_{55} =
 H_3^{-3/2}H_5^{-1}
  (H_3^{1/2} \sin^2 \theta + H_5 H_3^{-1/2} \cos^2 \theta)
$$
Now to calculate the potential as seen
by the M5 brane we put $H_3 = 1$. This reduces to
\be\label{sqdetg}
\sqrt{\det G} = (H_5^{-1}\sin^2\theta +  \cos^2\theta)^{1/2}.
\ee
As discussed earlier one can trust supergravity solution if 
$C\ll 1$. Therefore we have the following limits:

\begin{itemize}
\item
For $r \rightarrow\infty$
all the harmonic functions become $1$ and so $\sqrt{\det G} =1$. 

\item
Near $r \rightarrow 0$, $H_5 \rightarrow R/r$ and we can neglect 
$\sin^2\theta$. This gives
$$
\sqrt{\det G} =  {1\over \sqrt{1 + C^2}}.
$$
This is precisely the reduction expected from the BPS analysis in the 
previous sections.

\end{itemize}

\subsection{Lorentz invariance}
The origin $r=0$ is a smooth point for the metric (\ref{smrmet}).
If we rescale:
$$
\widetilde{x}_1 \equiv \sqrt{1+C^2} x_1,
$$
then at the vicinity of $r=0$ the $SO(5,1)$ Lorentz invariance
is restored. This is the same rescaling found in section (\ref{bps}).
As we will see in section (\ref{decoup}), the low-energy theory
that describes the M5-branes breaks $SO(5,1)$ Lorentz invariance,
but at lowest order the breaking term involves only the fermions.

Let us also mention that even though the metric is smooth, the
4-form field strength, $dC_3$, is discontinuous at $r=0$.
It is nevertheless finite and that will play an important role
in generating the Lorentz-breaking fermion term
(see section (\ref{loweng})).

\subsection{Small fluctuations}
We can expand the potential (\ref{sqdetg}) in small $r$.
To leading order,
$$
\sqrt{\det G} = 
 {1\over \sqrt{1 + C^2}} + {{C^2}\over {2 (1+C^2)^{1/2}}}{r\over R}
 + O(r)^2.
$$
If we change coordinates to $u = r^{1/2}$ and rescale $x_1$
by the factor $\sqrt{1+C^2}$, we see that fluctuations 
of the position of the M5-brane have an effective potential of
$$
{{C^2}\over {2(1+C^2)}} M_p u^2.
$$
We conclude that the fluctuations become massive with the same mass
as predicted by the BPS calculation (\ref{bpsmass}).


\section{Decoupling limits}\label{decoup}
We now wish to study limits of the previous constructions where
gravity can be decoupled.
To be concrete, we will concentrate on the example of $N$ M5-branes
at the center of a Taub-NUT space with $C_{167}$-flux turned on.

There are two kinds of decoupling arguments that we can utilize.
The first is a low-energy argument where we set to infinity the scale of 
all the excitations that we wish to discard.
Thus, we say that $\SUSY{4}$ SYM theory describes $N$ coincident
D3-branes when we set the string-scale to infinity, thereby decoupling
the massive string states \cite{WitBND}.
We will call this type of argument ``kinematical''.
Sometimes we are forced to keep the scale of excitations, that we wish
to decouple, finite.
Another type of decoupling argument is possible if we can set
the coupling constant between those excitations and the excitations
of our theory to zero.
This type of argument was introduced in \cite{SeiVBR}
for the decoupling of bulk string states from the little-string theory.
The bulk states have masses of order the string scale $M_s$ which is
the same scale as that of the little-string theory. The decoupling
is argued to occur in the limit of zero string coupling constant.
We will call this a ``dynamical'' argument.

\subsection{Kinematical decoupling}
We have seen in section (\ref{bps}) that an M5-brane at the center
of a Taub-NUT space with the field $C_{167}$ turned on has massive
BPS states. Let us make a list of the various energy scales that
we found in section (\ref{bps}).
In the limit of $C\rightarrow 0$, we have:
\begin{itemize}
\item
$M_p$ is the Planck scale.
\item
$R_7^{-1}$ is the scale of KK-excitations far away from the center.
\item
$M_p C^{1/3}$ is the energy-scale of the 
binding energy per unit volume of the 5-brane.
\item
$C R_7^{-1}$ is the energy scale of excitations of the 5-brane.
\end{itemize}

In a low-energy decoupling limit we must set the first three scales 
to infinity. The Planck scale, $M_p$ must be set to infinity
in order to decouple gravity.
The scale $M_p C^{1/3}$ must be set to infinity so that we will not
have to consider local fluctuations where a small portion
of the  M5-brane escapes to infinity.
Finally, the scale $R_7^{-1}$ will be sent to infinity so as to keep
the scale of Kaluza-Klein particles that are far from the center
small.
We will keep $C R_7^{-1}$ finite.
Finally, we should also take $M_p R_7\rightarrow\infty$.
The reason is as follows.
If we reduce from M-theory to type-IIA along $R_7$, the Taub-NUT
space becomes a D6-brane and the M5-branes becomes NS5-branes.
The string scale is $M_s^2 = M_p^3 R_7$ and this must be set
to infinity.\footnote{We thank S. Sethi for pointing this out.}
Moreover, we want the scale that is set by the tension of the
D6-brane to be much higher than the scale set by the tension
of the M5-branes, otherwise we could not decouple the 
6+1D $U(1)$ gauge field on the D6-brane.
The implies that the string coupling constant should be large,
and hence $M_p R_7\rightarrow\infty$.

The decoupled theory that we obtain 
seems to be a new type of 5+1D theory that has not been
encountered before.

\subsection{Dynamical decoupling}
In this limit we take $M_p\rightarrow\infty$ but we wish to keep
$C$ finite. Since we wish to keep $C R_7^{-1}$ finite as well,
we are forced to keep the scale $R_7^{-1}$ itself finite.
This means that we must find another argument for the decoupling of
Kaluza-Klein excitations that are far from the center.
Such an argument has to be dynamical, namely that the coupling constant
between these states and the states of the 5+1D theory is proportional
to inverse powers of $M_p$. This argument is similar in spirit to
the argument made in \cite{SeiVBR} for
the decoupling of bulk string states from the little-string theory.
The bulk states have masses of order the string scale $M_s$ which is
the same scale as that of the little-string theory. The decoupling
is argued to occur in the limit of zero string coupling constant.

In our case, in the limit of keeping both $R_7$ and $C$ finite,
we saw in section (\ref{bps}) that the metric in the $1^{st}$
direction has to be rescaled by a factor of $\sqrt{1+C^2}$
in order to preserve Lorentz invariance, at least to leading 
order.
By this we mean that the energy of massless particles with
low momentum $p$, will be $E=|p|$ no matter what the direction
of $p$ is. We have also seen that with this rescaling
the energy of tensor fluxes in directions
$2,3$ (i.e. not including the direction of the $C$-flux) is:
$$
E_{23} =   {{N_{23}^2 R_2 R_3}\over {2 N \wR_1 R_4 R_5}},
$$
where $N_{23}$ is an integer.
Fluxes in directions $1,2$ have energy:
$$
E_{12} = 
  {{N_{12}^2 \wR_1 R_2}\over {2 N R_3 R_4 R_5}}.
$$
which also respects the invariance under interchange of directions
$1$ and $2$.

At higher orders in the momentum expansion,
we expect Lorentz invariance  to be broken.
As we shall see in section (\ref{loweng}), it is the fermions
that first exhibit the breakdown of Lorentz invariance.

\section{Low-energy description}\label{loweng}
What is the low-energy description of these 5+1D theories?
A regular M5-brane is described, at low-energies, by a tensor
multiplet of $\SUSY{(2,0)}$ supersymmetry. It contains
an anti-self-dual tensor field and 5 scalars.
In our case, we have argued that the M5-brane becomes pinned
in 4 out of the 5 transverse directions.
Thus, it is reasonable to assume
that 4 scalars become massive and the low-energy
description is an anti-self-dual tensor field and a single scalar.
Together with fermions, that would make up a single tensor
multiplet of $\SUSY{(1,0)}$ supersymmetry and the equations of motion
would be Lorentz invariant at the lowest order in the derivative
expansion.

\subsection{The fermions: a puzzle}
What about the fermions?
At first sight, there seems to be a puzzle.
We have seen that 4 scalars become massive.
In terms of $\SUSY{(1,0)}$ the 4 scalars are part of a hyper-multiplet.
However, a hypermultiplet in 5+1D contains chiral fermions and
these cannot be made massive.
Furthermore, imagine that we turn on $C$ gradually from
$0$ to its present value. The theory at $C = 0$ has a (massless)
hypermultiplet that contributes to the gravitational anomaly.
How did this part of the anomaly disappear at $C\neq 0$?

The resolution of both puzzles is that the theory at $C\neq 0$
is {\em not Lorentz invariant}.
Thus, we should only consider the $SO(4,1)\subset SO(5,1)$ subgroup
of Lorentz-invariance and therefore only 4+1D supersymmetry.
In 4+1D, a hypermultiplet can be given a mass.

Let $\lam$ be the fermion that lives on an M5-brane.
It is in the representation $(\rep{4},\rep{4})$
of $Spin(4,1)\times Spin(5)$ where $Spin(5)$ is the rotation
of the transverse directions.

$Spin(4,1)$ invariance in directions $0,2\dots 5$ suggests
that the effective coupling would be:
\be\label{omegal}
\sim {\bar\lam} \Gamma^{1} \Omega\lam.
\ee
Here $\Gamma^1$ is one of the 5+1D Dirac matrices and 
$\Omega$ acts on the $Spin(5)$ R-symmetry indices.
Furthermore, $Spin(3)$ invariance in directions $8,9,10$
suggests that $\Omega$ should be a constant $Spin(3)\subset Spin(5)$
invariant matrix.
After dimensional reduction to 4+1D the term must reduce to
an ordinary mass term for the fermions of the hypermultiplet
because the theory would be Lorentz invariant.
This means that $\Gamma^{1}\Omega$ must be the identity
on the fermions of the hypermultiplet.
Starting with a vector multiplet with 16 supersymmetries in
4+1D we can consider an $\SUSY{1}$ subgroup ($8$ supersymmetries)
of the SUSY algebra.
The fermions that go into the hypermultiplet can be separated
from those that go into the vector multiplet (in 4+1D)
by their transformation under the $Spin(4)$ that, in the original
setting, corresponds to rotations in directions $7,8,9,10$.
The fermions of the hypermultiplet are invariant and this determines
$\Omega = \Gamma^6$ -- the 11D Dirac matrix in the $6^{th}$ direction.
Note that we cannot set $\Omega$ to the identity because 
then the effective action will be just like a coupling to a constant
gauge field that can be gauged away.

\subsection{The supergravity mechanism}
What is the explicit mechanism by which our fermions get a mass?
If one places an M5-brane in a region where the 4-form field
strength, $F$, of M-theory is nonzero but still small,
the fermions, $\lam$, on the world-volume of an M5-brane
(that are part of the tensor multiplet) couple to it schematically
as $\lam\lam F$.
One way to see this is by expanding the action with the 
fermionic zero-modes of the M5-brane solution.
Recall that M-theory has a term in the effective action
that is of the form \cite{CJ} ${\bar\psi^M}\Gamma^{PQ}\psi^N F_{MPQN}$,
where $M,P,Q,N=0\dots 10$, $\psi^M$ is the gravitino field,
$F_{MPQN}$ is the 4-form field-strength and $\Gamma^{PQ}$ is an
anti-symmetric product of two Dirac matrices.
After reduction to the zero-modes on the M5-brane solution
we can find  the effective coupling between the 5+1D
fermions and the external 4-form field-strength.
(Note that the M5-brane solution has strong curvature
but we can still use it for this discussion since supersymmetry
determines that term uniquely.)

In our case, let $\mu=0\dots 5$ a direction parallel to the M5-brane.
Let $A=6\dots 10$ be a direction orthogonal to the M5-brane.
We are interested in the coupling between the components
$F_{\mu ABC}$ of the field-strength and the fermions $\lam$
from the $(2,0)$ tensor-multiplet.
This coupling has to be of the form
$F_{\u ABC}{\bar\lam} \Gamma^\mu \Gamma^{ABC}\lam$, where $\Gamma^{ABC}$
acts on the $Sp(2)$ R-symmetry indices only and $\Gamma^\mu$
acts on the space-time $Spin(5,1)$ spinor indices only.

We have seen in section (\ref{smallr}) that the 4-form 
field-strength, $F_4$,  approaches a constant magnitude as
$r\rightarrow 0$ but it is still discontinous because its
direction depends on the path along the transverse
directions ($7,8,9,10$) in which we take the limit $r\rightarrow 0$.
At first sight this would seem to suggest that the $F \lambda\lambda$
term has a scalar-field dependent coefficient.
We believe, however, that the correct procedure is to expand
the gravitino fields in the presence of the M5-brane as 
$\psi^M \sim \lambda \psi^M_0$, where $\psi^M_0$ are
the gravitino zero-modes and the $\lambda$'s depend only on
directions $0\dots 5$ (along the M5-brane).
Then, we have to plug this back into the M-theory coupling,
${\bar\psi^M}\Gamma^{PQ}\psi^N F_{MPQN}$, with $F_{MNPQ}$ taken from
(\ref{Cthree}). We believe that this will produce the 
term suggested in (\ref{omegal}).

We could be a bit more precise here. Observe that the value of $F_{r167}$
near $r\to 0$ is a constant given by $C{(1 + C^2)R^{-1}}$. 
Therefore we expect
the fermions to pick up a mass proportional to the value of $F$ at the 
origin. However this is not the case. It is of course true that the 
$SO(4,1)$-invariant
 mass involves $F$ but there is also a contribution from the
zero modes of the gravitino in the picture. In the 
presence of the background $C$ field,
the normalisable zero modes also pick up 
contributions from the $C$ field in such a way that the zero modes are 
actually suppressed by inverse powers of $(1+C^2)$.
To see this, consider the D5-NS5 brane configuration.
Let us assume that the
system supports a normalisable gravitino zero mode $\psi^7_0(r)$.
We now go to the slanted torus by the transformation given
in section (\ref{grasol}) and then make
a T-duality along $x_7$. Using the T-duality
rules the combined effect now gives
a zero mode suppressed by $(1+C^2)^{-1/2}$.
Now integrating out the zero modes using the $11$-dimensional
term (with $x_1$ being scaled by a factor of $\sqrt{1+C^2}$)
we see that after dimensional reduction to 4+1D
the fermions get the same mass as the bosons.


\section{Various BPS states in 6+1D}\label{sevend}
We will now drop the M5-branes from the story and concentrate only
on the Taub-NUT space with the C-flux.
For the purposes of the discussion it is convenient to think of it
as a 6+1D theory, although we do not necessarily claim that it is
decoupled from gravity. We will return to that question in section
(\ref{largeC}).

We can calculate the tension of various BPS states of this 6+1D ``theory''.
We can do this using the same techniques as described in
section (\ref{bps}). We can even check some of
the statements for small $C$ using the solution
in section (\ref{grasol}).
As in section (\ref{bps}) we will assume that the theory
is compactified on $\MT{6}$ with radii $R_1,\dots, R_6$.
We will express the results in terms of the rescaled radii:
$$
\wR_1\equiv \sqrt{1+C^2} R_1,\qquad
\wR_6\equiv \sqrt{1+C^2} R_6.
$$
We have  calculated the energy of the following BPS objects:
\begin{itemize}
\item {\bf Kaluza-Klein particles:}
They have energy:
$$
\sqrt{ {{k_1^2}\over {\wR_1^2}} 
 + {{k_2^2}\over {R_2^2}}
 + \cdots 
 + {{k_5^2}\over {R_5^2}}
 + {{k_6^2}\over {\wR_6^2}}}.
$$

\item {\bf M2-branes:}
M2-branes that are stretched in directions $I\neq J$ have the
following masses, according to the dimension of
the intersection of the plane of the membrane with
the plane of the C-flux.
  \begin{itemize}
  \item
  For $I,J=2,3,4,5$ we have the mass:
  $$
  {{1}\over {\sqrt{1+C^2}}} M_p^3 R_I R_J.
  $$

  \item
  For $I,J$ with $I=1,6$ and
  $J=2,3,4,5$ we find the mass:
  $$
  {{1}\over {\sqrt{1+C^2}}} M_p^3 \wR_I R_J
  $$

  \item
  The BPS formula for M2-branes in direction $1,6$ gives:
  $$
  {1\over {\sqrt{1+C^2}}} M_p^3 \wR_1\wR_6,
  $$
  which again agrees with the supergravity calculation
  as in (\ref{grasol}), for small $C$.

  \end{itemize}

\item {\bf M5-branes:}
This again depends on whether the M5-brane hyper-plane contains
both the $1^{st}$ and $6^{th}$ directions or just one of them.

  \begin{itemize}
  \item
  For M5-branes in direction $1\dots 5$ we find the mass:
  $$
  {1\over {1+C^2}} M_p^6 \wR_1 R_2 R_3 R_4 R_5.
  $$

  \item
  For M5-branes in direction $1,3\dots 6$ we find the mass:
  $$
  {1\over {1+C^2}} M_p^6 \wR_1 R_3 R_4 R_5 \wR_6.
  $$
  \end{itemize}

\item
{\bf Electric fluxes}:
For this purpose we can think of the Taub-NUT as a D6-brane
(after reduction on the $7^{th}$ direction).
  \begin{itemize}
  \item
  For electric flux in the $1^{st}$ direction we find the central charge:
  $$
  Z = M_p^9 V R_7 \Gamma^{78} -i C_{167} R_1 R_7 \Gamma^{68}
    + M_p^3 R_1 R_7\Gamma^{17}
    + i C_{167} M_p^6 V R_7\Gamma^{16}
  $$
  and the energy is:
  $$
   {{\sqrt{1+C^2} \wR_1}\over {2 M_p^3 R_2 R_3 R_4 R_5 \wR_6}}.
  $$

  \item
  For flux in the $2^{nd}$ direction we find:
  $$
  Z = M_p^9 V R_7 \Gamma^{78}
    + M_p^3 R_2 R_7\Gamma^{27}
    + i C_{167} M_p^6 V R_7\Gamma^{16}
  $$
  The energy is:
  $$
  {{\sqrt{1+C^2} R_2}\over {2 M_p^3 \wR_1 R_3 R_4 R_5 \wR_6}}.
  $$
  \end{itemize}

\item {\bf Magnetic fluxes:}
We take the magnetic flux to be in direction $I,J$.
We will distinguish three cases:
  \begin{itemize}
  \item
  For $I,J=2,\dots 5$, for example $I,J=2,3$, we find that the central charge is:
  $$
  Z = M_p^9 V R_7 \Gamma^{78}
    + i M_p^3 R_1 R_4 R_5 R_6 R_7\Gamma^{23}
    - C_{167} R_1 R_4 R_5 R_6 R_7\Gamma^{45}
    + i C_{167} M_p^6 V R_7\Gamma^{16},
  $$
  and the energy is:
  $$
  {{M_p^3 \wR_1 R_4 R_5 \wR_6}\over {2\sqrt{1+C^2} R_2 R_3}}.
  $$

  \item
  For $I=2,\dots 5$ and $J=1,6$, for example, $I=2$ and $J=1$,
  we find:
  $$
  Z = M_p^9 V R_7 \Gamma^{78}
    + i M_p^3 R_3 R_4 R_5 R_6 R_7\Gamma^{12}
    + i C_{167} M_p^6 V R_7\Gamma^{16},
  $$
  and the energy is:
  $$
  {{M_p^3 R_3 R_4 R_5 \wR_6}\over {2\sqrt{1+C^2}\wR_1 R_2}}
  $$

  \item
  For $I,J=1,6$, we find the energy:
  $$
  {{C}\over {\sqrt{1+C^2}}} M_p^6 R_2 R_3 R_4 R_5 R_7.
  $$
  \end{itemize}
\end{itemize}

\subsection{The large $C$ limit}\label{largeC}
Perhaps the most interesting limit to study is that
of $C\rightarrow\infty$.\footnote{We wish to thank S. Sethi
for pointing this limit out.}
Specifically, let us consider the following limit:
\be\label{ourlim}
M_p\rightarrow\infty,\,\,
C\rightarrow\infty,\,\,
M_p^3 C^{-1}\rightarrow {\mbox{fixed}}.
\ee
This limit has been studied in \cite{AOSJ}.
They argued that a D6-brane
in the limit of noncommutative geometry \cite{CDS}-\cite{SWNCG}
is described by a decoupled 6+1D theory.
Keeping $M_p^3 C^{-1}$ finite makes sures that the effective 
Yang-Mills coupling constant $g_{YM}^2 = M_p^{-3} C$
is fixed.
The proposal of a decoupled theory is also related to
a previous suggestion of \cite{ABS}
that a non-commutative version of the 6+1D theory that
should have been the M(atrix)-model of M-theory on $T^6$
could perhaps avert the problems explained in \cite{SeiWHY,SenWHY}
and actually be a decoupled theory.
We will not address the issue of decoupling in this paper.
(Following a correspondence with O. Aharony, we tend to believe
that the 6+1D theory is decoupled, as suggested in \cite{AOSJ},
but has a continuum of states like a 10+1D theory.)

Nevertheless, we will point out that all the energies of the BPS 
states studied in the previous section, except
the magnetic flux in directions $1,6$,
have a finite limit if $M_p^3 C^{-1}$ is kept fixed
and $R_7$ is also kept finite.

The mass of the M2-branes
is proportional to $1/g^2_{YM}$, the mass of the M5-branes is
proportional to $1/g^4_{YM}$ and the energy of electric-fluxes in
directions orthogonal to $1$ and $6$ is proportional to $g^2_{YM}$. 
It is interesting to note that the M2-branes and M5-branes
are becoming light when $g^2_{YM}$ is large (compared to the 
radii $\wR_1,R_2,\dots,R_5,\wR_6$). 


\section{Discussion}
We have seen that the dynamics near the origin of a Taub-NUT
space with a $C$-field turned on at $\infty$ can be used
to construct various decoupled theories.
We have suggested two types of theories.
These theories can be obtained by placing M5-branes
as probes and taking either a low-energy
limit or a less understood dynamical decoupling limit.
In all these examples, the dynamics depends only on the type
of singularity at $r=0$. The virtue of using a Taub-NUT space
is that we can easily use BPS arguments as in section (\ref{bps}).

We can repeat the discussions of the previous sections
with D-branes of various dimensions instead of M5-branes
or with M2-branes.
When we discuss D-branes we can turn on various RR-fields
or NSNS 2-form fields at infinity.
These fields can have various numbers of indices along the 
direction of the branes. Thus, we will obtain theories
that are either Lorentz invariant or have a
Lorentz-breaking term that is characterized by
a vector or a tensor.

Alternatively, we can compactify the 
5+1D theory that we found on the M5-branes on $\MT{d}$
and look for a low-energy description of the resulting
$(6-d)$-dimensional theory.
Since the Lorentz-breaking direction (the $1^{st}$ direction
in the notation of section (\ref{bps})) can be either compactified
or not, we obtain low-energy descriptions that are either
Lorentz-invariant or have a vector Lorentz-breaking term.

The two questions, one regarding replacing M5-branes with
D-branes or M2-branes and the other regarding  the low-energy
description of compactified theories, are overlapping.
For example, if we compactify the theory of ordinary M5-branes,
i.e. the $(2,0)$-theory, on $\MT{2}$ we obtain the theory
of D3-branes, i.e. $U(N)$ $\SUSY{4}$ SYM at low-energy.
If we further compactify the theory of D3-branes on $\MS{1}$
we obtain the theory of D2-branes that at the IR limit flows
to the $Spin(8)$
theory of M2-branes at a certain point in moduli space
\cite{SetSus}-\cite{SeiSXN}.

These theories seem to provide a mechanism for fixing the position
of branes in an ambient space.
It would be interesting to generalize these constructions
to  cases, with probably less
supersymmetry, in which more than 4 coordinates of the position
of a D-brane are fixed. This might have applications
for brane-world scenarios as in \cite{ADD}-\cite{RanSun}.

Let us also mention that our models are related to 
the models studied in \cite{JPP}.
Both models are U-dual to the elliptic brane configurations
of \cite{WitBR}.

Another question that we hope to address in a later paper
is the large $N$ limit of this theory, along the lines
of the AdS/CFT correspondence \cite{Juan,HasItz,MalRus}.

It would also be interesting to study the limit of large $C$
as discussed in (\ref{largeC}).



\section*{Acknowledgments}
It is a pleasure to  thank M. Berkooz, M.B. Green, A. Kapustin,
S. Mukhi, A. Sen and S. Sethi for helpful discussions.
We are also grateful to O. Aharony and T. Banks for helpful 
correspondence.
G.R. gratefully acknowledges the hospitality of the Mathematical Institute at
the University of Oxford, where part of this work was done.

The research of K.D. is supported by Department of Energy grant No. 
DE-FG02-90ER4054442.
The research of O.J.G was supported by National Science Foundation grant
No. PHY98-02484.
The research of G.R. is supported by NSF grant number NSF DMS-9627351.

\def\np#1#2#3{{\it Nucl.\ Phys.} {\bf B#1} (#2) #3}
\def\pl#1#2#3{{\it Phys.\ Lett.} {\bf B#1} (#2) #3}
\def\physrev#1#2#3{{\it Phys.\ Rev.\ Lett.} {\bf #1} (#2) #3}
\def\prd#1#2#3{{\it Phys.\ Rev.} {\bf D#1} (#2) #3}
\def\ap#1#2#3{{\it Ann.\ Phys.} {\bf #1} (#2) #3}
\def\ppt#1#2#3{{\it Phys.\ Rep.} {\bf #1} (#2) #3}
\def\rmp#1#2#3{{\it Rev.\ Mod.\ Phys.} {\bf #1} (#2) #3}
\def\cmp#1#2#3{{\it Comm.\ Math.\ Phys.} {\bf #1} (#2) #3}
\def\mpla#1#2#3{{\it Mod.\ Phys.\ Lett.} {\bf #1} (#2) #3}
\def\jhep#1#2#3{{\it JHEP.} {\bf #1} (#2) #3}
\def\atmp#1#2#3{{\it Adv.\ Theor.\ Math.\ Phys.} {\bf #1} (#2) #3}
\def\jgp#1#2#3{{\it J.\ Geom.\ Phys.} {\bf #1} (#2) #3}
\def\cqg#1#2#3{{\it Class.\ Quant.\ Grav.} {\bf #1} (#2) #3}
\def\hepth#1{{\it hep-th/{#1}}}
\def\hepph#1{{\it hep-ph/{#1}}}


\end{document}